# Single-photon-emitting optical centers in diamond fabricated upon Sn implantation

S. Ditalia Tchernij[1,2], T. Herzig[3], J. Forneris[2,1]*, J. Küpper[3], S. Pezzagna[3], P. Traina[4], E. Moreva[4], I.P. Degiovanni[4], G. Brida[4], N. Skukan[5], M. Genovese[4,2], M. Jakšić[5], J. Meijer[3], P. Olivero[1,2]

[1] Physics Department and "NIS" Inter-departmental Centre - University of Torino; via P. Giuria 1, 10125, Torino, Italy
[2] Istituto Nazionale di Fisica Nucleare (INFN) - Sez. Torino; via P. Giuria 1, 10125, Torino, Italy
[3] Department of Nuclear Solid State Physics; University of Leipzig, 04103, Leipzig, Germany
[4] Istituto Nazionale di Ricerca Metrologica (INRiM); Strada delle Cacce 91, 10135, Torino, Italy
[5] Ruđer Bošković Institute; Bijenicka 54, P.O. Box 180, 10002, Zagreb, Croatia

\* corresponding author. Email: forneris@to.infn.it

## Abstract

The fabrication of luminescent defects in single-crystal diamond upon Sn implantation and annealing is reported. The relevant spectral features of the optical centers (emission peaks at 593.5 nm, 620.3 nm, 630.7 nm and 646.7 nm) are attributed to Sn-related defects through the correlation of their photoluminescence (PL) intensity with the implantation fluence. Single Sn-related defects were identified and characterized through the acquisition of their second-order auto-correlation emission functions, by means of Hanbury-Brown & Twiss interferometry. The investigation of their single-photon emission regime as a function of excitation laser power revealed that Sn-related defects are based on three-level systems with a 6 ns radiative decay lifetime. In a fraction of the studied centers, the observation of a blinking PL emission is indicative of the existence of a dark state. Furthermore, absorption dependence from the polarization of the excitation radiation with ~45% contrast was measured. This work shed light on the existence of a new optical center associated with a group-IV impurity in diamond, with similar photo-physical properties to the already well-known Si-V and Ge-V emitters, thus providing results of interest from both the fundamental and applicative points of view.

# Single-photon-emitting optical centers in diamond fabricated upon Sn implantation

S. Ditalia Tchernij[1,2], T. Herzig[3], J. Forneris[2,1*], J. Küpper[3], S. Pezzagna[3], P. Traina[4], E. Moreva[4], I.P. Degiovanni[4], G. Brida[4], N. Skukan[5], M. Genovese[4,2], M. Jakšić[5], J. Meijer[3], P. Olivero[1,2]

[1] Physics Department and "NIS" Inter-departmental Centre - University of Torino; via P. Giuria 1, 10125, Torino, Italy
[2] Istituto Nazionale di Fisica Nucleare (INFN) - Sez. Torino; via P. Giuria 1, 10125, Torino, Italy
[3] Department of Nuclear Solid State Physics; University of Leipzig, 04103, Leipzig, Germany
[4] Istituto Nazionale di Ricerca Metrologica (INRiM); Strada delle Cacce 91, 10135, Torino, Italy
[5] Ruđer Bošković Institute; Bijenicka 54, P.O. Box 180, 10002, Zagreb, Croatia

* corresponding author. Email: forneris@to.infn.it

## Abstract

The fabrication of luminescent defects in single-crystal diamond upon Sn implantation and annealing is reported. The relevant spectral features of the optical centers (emission peaks at 593.5 nm, 620.3 nm, 630.7 nm and 646.7 nm) are attributed to Sn-related defects through the correlation of their photoluminescence (PL) intensity with the implantation fluence. Single Sn-related defects were identified and characterized through the acquisition of their second-order auto-correlation emission functions, by means of Hanbury-Brown & Twiss interferometry. The investigation of their single-photon emission regime as a function of excitation laser power revealed that Sn-related defects are based on three-level systems with a 6 ns radiative decay lifetime. In a fraction of the studied centers, the observation of a blinking PL emission is indicative of the existence of a dark state. Furthermore, absorption dependence from the polarization of the excitation radiation with ~45% contrast was measured. This work shed light on the existence of a new optical center associated with a group-IV impurity in diamond, with similar photo-physical properties to the already well-known Si-V and Ge-V emitters, thus providing results of interest from both the fundamental and applicative points of view.

## Introduction

Single-photon sources (SPS) are essential building blocks for the development of quantum technologies, ranging from fundamental quantum optics experiments to quantum metrology and quantum key distribution [1-4]. In recent years, a large number of materials and systems have been explored with the purpose of identifying reliable SPS, such as quantum dots [5,6], 2-dimensional [7] and wide bandgap materials [8-11]. Among these systems, diamond is an appealing material, as its lattice can host different defects with bright and photostable single-photon emission at room temperature. The quest for single-photon emitters with desirable opto-physical properties has led to the discovery and characterization of several classes of optical centers in diamond, from the nitrogen-vacancy complex (NV center) [12] to alternative emitters based on Si [13,14] and Ge [15] impurities, as well as additional centers in the visible [16,17] and near-infrared [18-20] spectral range.

Since Si-V and Ge-V centers are among the most appealing optical centers in diamond for applications as SPS, due to their photo-stability and short emission lifetime, the existence of additional optically active defects associated with group IV impurities in diamond and the assessment of their opto-physical properties would be of high interest, from both fundamental and applicative points of view.

Here we report for the first time, to the best of our knowledge, on the evidence of Sn-related single-photon emitters in single-crystal diamond fabricated upon ion implantation and subsequent annealing. In particular, the PL characterization of the peculiar emission features is performed for different ion implantation fluences to unambiguously attribute them to Sn-related defects. Their emission lifetime is evaluated from the acquisition of the second-order auto-correlation function of individual emitters. The photo-stability and the polarized absorption of the centers are also discussed on the basis of the experimental results at the single-photon emitter level.

## 2. Experimental

The measurements were performed on a 2×2×0.3 mm³ single-crystal diamond substrate produced by ElementSix. The sample was denoted as "electronic grade" due to the low nominal concentration of substitutional N and B (<5 ppb and <1 ppb, respectively). Several ~200×200 μm² regions were implanted at different Sn energies and fluences. One region underwent a 10 MeV Sn implantation at 5×10¹³ cm⁻² fluence at the Laboratory for Ion Beam Interaction of the Ruđer Bošković Institute. Three additional regions were implanted with 60 keV Sn ions in the 3-10×10¹¹ cm⁻² fluence range at the low-energy accelerator of the University of Leipzig. At the same facility, one additional region was implanted with 60 keV Sn at lower (i.e. 1×10¹¹ cm⁻²) fluence to allow the investigation of individual Sn-related centers. The substrate was subsequently annealed for 2 hours in vacuum at 950 °C to promote the formation of optically active defects. A subsequent oxygen plasma cleaning (pressure: 0.3 mbar, gas flow: 8 sccm, power: 200 W) was performed to minimize the surface fluorescence associated with the sample graphitization and contamination occurred during the thermal treatment.

## 3. Results

*PL characterization at ensemble level.* The ensemble PL emission from the regions implanted at 3×10¹¹ – 5×10¹³ cm⁻² fluences was investigated with a Horiba Jobin Yvon HR800 Raman spectrometer (600 grooves mm⁻¹ diffraction grating, ~0.1 nm spectral resolution [21]). The excitation radiation was delivered by a cw 532 nm laser focused on the sample surface with a 20× air objective (21.6 mW power on the sample surface). The size of the probed spot region was ~3 μm, both in diameter and focal depth. The typical measured PL features are reported in the spectra shown in **Fig. 1a** and **Fig. 1b**, which were acquired from the region irradiated at 5×10¹³ cm² fluence. In particular, **Fig. 1b** shows the PL spectrum acquired under the same conditions with 514 nm laser excitation, to allow for an unambiguous attribution of the PL emission with respect to Raman emission. The PL spectrum consists of a weak emission line at 593.5 nm, with an estimated FWHM of 3 nm, followed by a bright PL peak at 620.3 nm (FWHM: 7 nm) and two less intense emission lines at 630.7 nm (FWHM: 5 nm) and 646.7 nm (FWHM: 2 nm). In the present study it was not possible to identify which of the observed PL peaks could be unequivocally attributed either to zero-phonon lines (ZPLs) associated with different charge states, or to their respective phonon sidebands. It is however worth noting that, if the 620.3 nm line (i.e. the most intense spectral feature) is considered as a ZPL, the following 630.7 nm peak would correspond to a 33 meV energy shift. The resonance frequency for quasi-local vibrations of a bound Sn atom in the diamond lattice is given by [22, 23]:

$$\Delta E = [m/3(M - m)]^{1/2} \, \omega_D \qquad (1)$$

where $m = 12$ and $M = 119$ are the atomic masses of C and Sn atoms and $\omega_D = 150$ meV is the Debye frequency of diamond. The calculated value is therefore $\Delta E \sim 29$ meV, a value in line with the above-mentioned shift.

Our attribution of the observed PL emission to Sn-related centers is based on the systematic dependence of the intensity of the reported emission lines from the ion implantation fluence. The PL spectra acquired at increasing fluences in the $3-10\times10^{11}$ cm$^{-2}$ range are reported together with a reference spectrum from an unirradiated region of the sample in **Figs. 1c-f**, after normalization to the intensity of the first-order Raman line (measured at 572.5 nm and not shown here). The emission lines at 593.5 nm, 620.3 nm, 630.7 nm and 646.7 nm are entirely absent from the unirradiated region (**Fig. 1c**). The only recurrent feature among the reported spectra is the second-order Raman emission of diamond, having characteristic peaks at 612.5 nm and 620.2 nm, corresponding respectively to 2470 cm$^{-1}$ and 2673 cm$^{-1}$ Raman shifts [24]. It is worth noting that the second-order Raman features of constant intensities were observed only in this set of measurements, due to the intense power of the laser source (i.e. 21.6 mW), and did not affect the characterization of single-photon emitters performed in the following at significantly weaker PL excitation (i.e. <3 mW). Conversely, the afore-mentioned PL emission lines increase in their intensity at increasing implantation fluences, up to the point where the Raman emission becomes negligible (**Figs. 1a,b**). Notably, the 620.3 nm PL peak strongly overlaps with the 2673 cm$^{-1}$ Raman peak, therefore its intensity needed to be suitably deconvoluted from the Raman background. As mentioned above, to provide an unequivocal attribution of the observed emission to PL and Raman transitions, the PL measurements were repeated at 514 nm excitation. In this case, the 2470 cm$^{-1}$ and 2673 cm$^{-1}$ Raman peaks shift to 588.7 nm and 595.9 nm, this time overlapping with the 593.5 nm PL emission line. However (see **Fig. 1b**), as expected all of the lines previously attributed to PL emission remain unchanged. The fact that these PL spectral features exhibit a strong correlation with the concentration of Sn in the diamond substrate, combined with their absence in the extensive body of PL characterization works carried on diamond implanted with different ion species [19,21,22,25], strongly supports their attribution to Sn-related complexes.

Despite a conclusive attribution of the defect nature is well beyond the scope of the present work, it is worth noting that the emission of Sn-related centers could be tentatively ascribed to a Sn-vacancy complex, in analogy with the well understood models for the other optical centers associated with group IV impurities (i.e. Si, Ge) in the diamond lattice [15, 26-28].

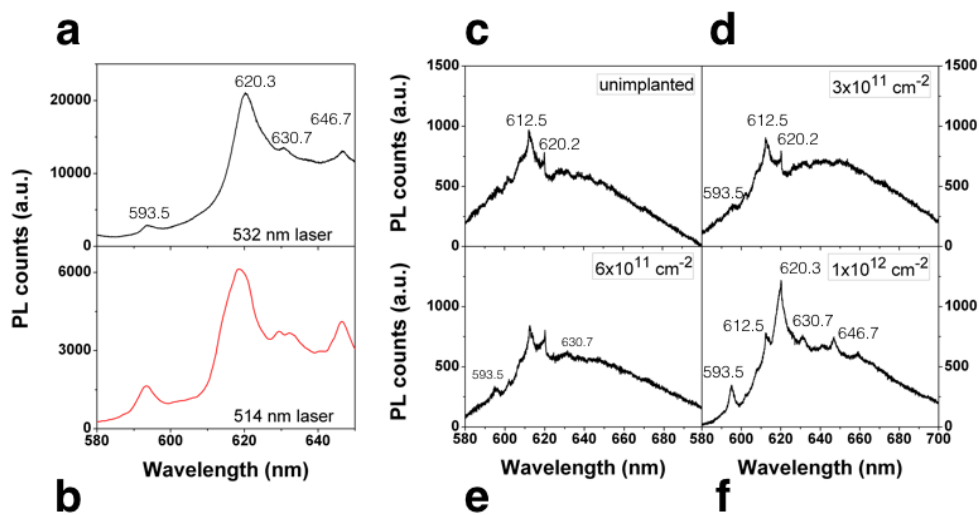

**Figure 1:** PL characterization of Sn-related centers upon ion implantation. PL spectra acquired under **a)** 532 nm and **b)** 514 nm CW excitations from the diamond region implanted with 10 MeV Sn ions at $5\times10^{13}$ cm$^{-2}$ fluence. **c)** reference spectrum from an unirradiated area. PL spectra acquired from different regions implanted with 60 keV Sn ions, at fluences of **d)** $3\times10^{11}$ cm$^{-2}$, **e)** $6\times10^{11}$ cm$^{-2}$ and **f)** $1\times10^{12}$ cm$^{-2}$.

*PL characterization at single-center level.* PL measurements were performed with a home-built single-photon-sensitive confocal microscope equipped with a 60× oil objective (1.35 numerical aperture), a 532 nm cw excitation from a ND:YAG laser, and two silicon avalanche photodiodes (SPADs). A suitable set of filters enabled to collect photons in the 580-638 nm spectral range. A Hanbury-Brown & Twiss interferometric configuration was exploited for the study of the second-order auto-correlation function from individual centers. The spectral analysis was carried out with an Oxford Instruments Mono-CL monochromator (1200 grooves mm$^{-1}$, 5 nm spectral resolution) which was fiber-coupled to a SPAD; in this case, the measurements were performed with a different set of filters, enabling the investigation in whole visible range at wavelengths $\lambda > 570$ nm.

**Fig. 2a** shows a 5×5 µm$^2$ PL confocal map acquired at room temperature from edge of the region implanted at $1\times10^{11}$ cm$^{-2}$ fluence. The map displays a density of isolated spots of about 1 spot per µm$^2$, all exhibiting similar confocal characteristics in terms of both spectral features and non-classical emission regime. A typical PL spectrum, acquired from an individual spot (highlighted in **Fig. 2a** with a circle) and subtracted by the PL spectrum obtained from an unimplanted region of the sample is shown in **Fig. 2b**. Within the limited spectral sensitivity of this system, the only apparent spectral feature is the most intense 620.3 nm line.

An investigation on the non-classical emission properties of isolated Sn-related centers was performed by acquiring their second-order auto-correlation PL chronograms $g^{(2)}(t)$ [4,29]. The expected second-order emission chronogram from the same single-photon emitter reported in Fig. 2 is shown in **Fig. 3a**. $g^{(2)}(t)$ chronograms were acquired at increasing laser excitation power, ranging from 80 µW to 720 µW.

The $g^{(2)}(t)$ chronograms, upon normalization according to the procedure described in Ref. [30], display an anti-bunching behavior, i.e. a pronounced dip at null delay time, indicating the emission of non-classical light from the center. More specifically, the $g^2(0) = 0.29 \pm 0.02$ value of the second-order auto-correlation function, estimated under 80 µW excitation power even before applying any correction for the negligible background luminescence, clearly indicates that the emission comes from an individual color center.

The Sn-related defects exhibited a bunching effect at high excitation powers ($P > 240$ µW), thus revealing the presence of a shelving state in the involved electronic transitions, similarly to what was reported for the NV, NE8 and NIR centers [12,18,19], as well as, more pertinently to this work, the group IV-related Si-V and Ge-V centers in diamond [15,31]. After suitable background subtraction [30,32], the PL emission lifetimes of the center was estimated by fitting the $g^{(2)}(t)$ curves with the following function describing a three-levels system [29]:

$$g^{(2)}(t)=1-(1-a_1)\cdot\exp(-|t|/\tau_1)+a_2\exp(-|t|/\tau_2) \qquad (2)$$

where $\tau_1$ and $\tau_2$ represent the characteristic times associated respectively to the anti-bunching and the bunching components of the chronogram. After having measured the $\tau_1$ parameter at different excitation powers from the same center (**Fig. 3a**), the emission center lifetime was estimated as the intercept of the linear fit of the trend of $\tau_1$ against the excitation power [19,33,34] as $\tau_1 = (6.0 \pm 0.1)$ ns (see **Fig. 3b**). This value is shorter than that of the NV center in bulk diamond (~12 ns, [35,36]), and in line with the lifetime of the Si-V and Ge-V centers in diamond (~2 ns and ~5 ns, respectively [13,15]). From the fitting procedure, the $\tau_2$ decay time was estimated as (68 ± 3) ns and (57 ± 2) ns for the 470 μW and 720 μW excitation powers, respectively.

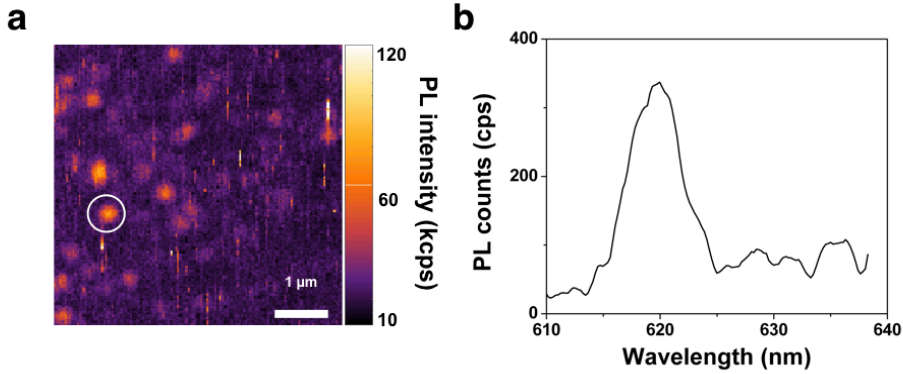

**Figure 2**: Sn-related single-photon emitters under 532 nm cw laser excitation. **a)** Confocal PL map of the sample region implanted with 60 keV Sn ions at the lowest fluence (1×10¹¹ cm⁻²). **b)** background-subtracted PL spectrum acquired from the individual Sn-related emitter highlighted in a) by the white circle.

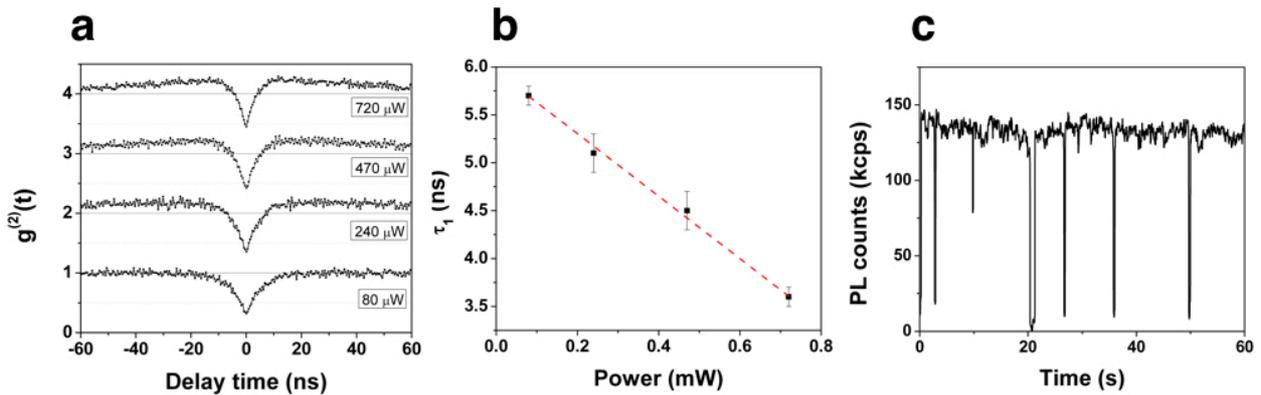

**Figure 3**: Non-classical PL emission properties of Sn-related centers. **a)** Second-order auto-correlation chronograms taken from the luminescent spot circled in white in Fig. 2a under increasing excitation powers in the 80-720 μW range. The curves are displaced along the vertical axis of unitary offsets, for sake of readability. **b)** Linear fit of the lifetime value $\tau_1$ extracted from a) as a function of the laser excitation power. The intercept value defines the emission lifetime. **c)** Temporal evolution of the PL emission rate of an individual Sn-related emitter displaying blinking behavior.

*Emission blinking.* The presence of a shelving state in the second-order auto-correlation function is accompanied by random photo-blinking over time, which affected a non-negligible fraction of the individual emitters investigated under intense 532 nm (i.e. >1 mW) laser excitation. A typical example of this phenomenon is reported in **Fig. 3c**, where the background-subtracted PL count rate is reported against the evolution time under continuous ~1 mW laser pump power. The blinking mostly consisted of an "on-off" behavior [16] between the full light emission from the center and the complete absence of Sn-related emission, and as expected was more pronounced at higher excitation powers. It is worth remarking that some centers permanently switched to a dark state upon exposure for ~15 s to >5 mW laser excitation. Differently from what reported for other single-photon emitters in diamond [19,37-39], no intermediate states with smaller decreases in the emission rates were observed in the investigations.

This observation can be tentatively attributed to the switch of the Sn defects to a "dark" charge state, possibly activated by non-radiative processes (e.g. resonant energy transfer) by surrounding defects [19,40], or by two-photon defect ionization induced by the intense laser excitation [41].

*Emission saturation.* **Fig. 4a** shows the typical dependence of the PL emission intensity from the laser excitation power, as acquired from an individual Sn-related center exhibiting photo-stable emission. The background-subtracted intensity trend follows a linear behavior at powers lower than 1 mW, followed by a saturation at higher powers. The trend was fitted with the following expression [13,18,31]:

$$I(P) = I_{sat} \cdot P/(P + P_{sat}), \tag{2}$$

where $I$ is the PL count rate and $P$ is laser excitation power. The saturation intensity and saturation excitation power were respectively evaluated as $I_{sat} = (1.37 \pm 0.01) \times 10^6$ photons s$^{-1}$ and $P_{sat} = (1.11 \pm 0.01)$ mW.

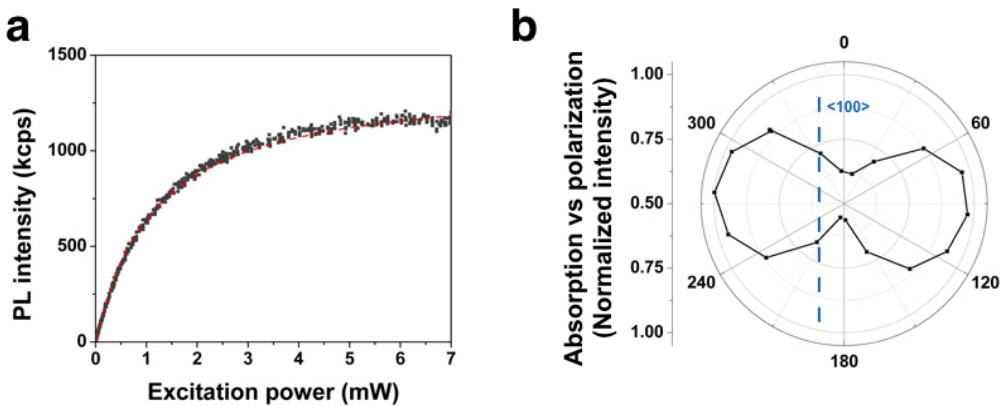

**Figure 4:** Emission properties of Sn-related centers. **a)** Emission rate of an individual Sn-related defect as a function of the laser excitation power. The red line represents the fitting curve. **b)** Intensity of the PL emission from an individual Sn-related center as a function of the polarization direction of the linearly polarized excitation laser with respect to the <100> crystallographic direction.

*Polarization measurements.* The dependence of the emission intensity from the polarization direction of the excitation beam was investigated by placing a linear polarizer at the excitation laser output (0.9 mW on the sample surface), followed by a half-wave plate, which was employed to rotate the excitation polarization. The normalized intensity of the background-subtracted PL emission acquired from an individual Sn-related center is reported in **Fig. 4b** as a function of the half-wave rotation angle. The <100> cyrstallographic direction of the sample corresponds to a $(0 \pm 5)°$ angle. A linearly polarized absorption is apparent, although with an extinction ratio of only ~45%. The absence of a fully polarized absorption is tentatively ascribed to the presence of several unfiltered Sn-related spectral features lines, whose polarization axes might not be aligned along the same direction, similarly to what is observed for the Si-V center [27].

## Conclusions

We reported on the controlled fabrication and preliminary photo-physical characterization of new single-photon emitters based on the incorporation of Sn impurities into the diamond crystal by direct ion implantation and subsequent annealing. The spectral features of the Sn centers emission consisted of peaks at 539.3 nm, 620.3 nm, 630.7 nm and 646.7 nm.

The Sn-related defects exhibit a three-level emission structure, with a radiative lifetime of 6 ns and a shelving state with characteristic decay times of 57-68 ns for laser pump powers in the 470 - 720 µW range, respectively. The presence of a shelving state is accompanied by emission blinking at high laser excitation powers, which is affecting a non-negligible fraction of the investigated centers. The Sn-related centers also displayed polarized light absorption with 45% extinction ratio, and a saturation PL emission of over 1 Mcps. These properties are comparable with those of the other known color centers related to group IV impurities in diamond, namely the Ge-V and Si-V centers. These results provide a significant contribution to the understanding and modeling of optically active centers based on the incorporation of group IV impurities in diamond, and open new perspectives into a novel class of bright single-photon emitters.

## Acknowledgements

This research activity was supported by the following projects: "DIESIS" project funded by the Italian National Institute of Nuclear Physics (INFN) - CSN5 within the "Young research grant" scheme; Coordinated Research Project "F11020" of the International Atomic Energy Agency (IAEA); EMPIR Project. No. 14IND05-MIQC2; NATO project G5263; Progetto Premiale 2014 "Q-SecGroundSpace" funded by MIUR. T.H, S.P. and J.M. gratefully acknowledge the support of Volkswagen Stiftung. J.F. gratefully acknowledges the CERIC-ERIC Consortium for the financial support to the access to the Laboratory for Ion Beam Interactions of the Ruđer Bošković Institute (proposal n. 20162021). S. D. gratefully acknowledges the "Erasmus Traineeship 2016-2017" program for the financial support to the access to the ion implantation facilities of the University of Leipzig.

# References

[1] M. Oxborrow M, A.G. Sinclair, *Single-photon sources*, Contemp. Phys. 46 (2005) 173. doi: 10.1080/00107510512331337936.

[2] M. D. Eisaman J. Fan, A. Migdall, S. V. Polyakov, *Invited Review Article: Single-photon sources and detectors*, Rev. Sci. Instr. 82, 071101 (2011). doi: 10.1063/1.3610677.

[3] C. J. Chunnilall, I. P. Degiovanni, S.Kück, I.Müller, A. G. Sinclair, *Metrology of single-photon sources and detectors: a review*, Opt. Eng. 53(8) (2014) 081910. doi: 10.1117/1.OE.53.8.081910.

[4] S. Prawer, I. Aharonovich, *Quantum Information Processing with Diamond: Principles and Applications.* ch. 6 (Woodhead Publishing, 2014).

[5] J. Claudon, J. Bleuse, N. Singh Malik, M. Bazin, P. Jaffrennou, N. Gregersen, C. Sauvan, P. Lalanne, J.-M. Gérard, *A highly efficient single-photon source based on a quantum dot in a photonic nanowire*, Nature Photonics 4, 174 - 177 (2010). doi: doi: 10.1038/nphoton.2009.287.

[6] A. Kiraz, M. Atatüre, A. Imamoğlu, *Quantum-dot single-photon sources: Prospects for applications in linear optics quantum-information processing*, Phys. Rev. A 69 (2004) 032305. doi: 10.1103/PhysRevA.69.032305.

[7] Jörg Wrachtrup, 2D materials: *Single photons at room temperature*, Nature Nanotechnology 11, 7–8 (2016). doi:10.1038/nnano.2015.265.

[8] I. Aharonovich, D. Englund, M. Toth, *Solid-state single-photon emitters*, Nature Photonics 10, 631–641 (2016) doi:10.1038/nphoton.2016.186.

[9] S. Pezzagna, D. Rogalla, D. Wildanger, J. Meijer, A. Zaitsev, *Creation and nature of optical centres in diamond for single-photon emission—overview and critical remarks*, New J. Phys. 13 (2011) 035024. doi: 10.1088/1367-2630/13/3/035024.

[10] A Lohrmann, B C Johnson, J C McCallum, S Castelletto, *A review on single photon sources in silicon carbide*, Rep. Prog. Phys. 80 (2017) 034502. doi: 10.1088/1361-6633/aa5171.

[11] S. Kako, C. Santori, K. Hoshino, S. Götzinger, Y. Yamamoto, Y. Arakawa, *A gallium nitride single-photon source operating at 200 K*, Nature Materials 5, 887 - 892 (2006). doi:10.1038/nmat1763.

[12] A. Beveratos, R. Brouri, T. Gacoin, J.-P. Poizat, P. Grangier, *Nonclassical radiation from diamond nanocrystals*, Phys. Rev. A 64 (2001) 061802. 10.1103/PhysRevA.64.061802.

[13] Wang C, Kurtsiefer C, Weinfurter H, Burchard B, *Single photon emission from SiV centres in diamond produced by ion implantation*, 2006 J. Phys. B: At. Mol. Opt. 39 37. doi: 10.1088/0953-4075/39/1/005.

[14] L. J. Rogers, K.D. Jahnke, T. Teraji, L. Marseglia, C. Müller, B. Naydenov, H. Schauffert, C. Kranz, J. Isoya, L.P. McGuinness, F. Jelezko, *Multiple intrinsically identical single-photon emitters in the solid state*, Nature. Comm. 5, 4739 (2014). doi: 10.1038/ncomms5739.

[15] T. Iwasaki, F. Ishibashi, M. Miyamoto, Y. Doi, S. Kobayashi, T. Miyazaki, K. Tahara, K. D. Jahnke, L. J. Rogers, B. Naydenov, F. Jelezko, S. Yamasaki, S. Nagamachi, T. Inubushi, N. Mizuochi, M. Hatano, *Germanium-Vacancy Single Color Centers in Diamond*, Scientific Reports 5 (2016) 12882. doi: 10.1038/srep12882.

[16] R. John, J. Lehnert, M. Mensing, D. Spemann, S. Pezzagna J. Meijer, *Bright optical centre in diamond with narrow, highly polarised and nearly phonon-free fluorescence at room temperature*, New J. Phys. 19 (2017) 053008. doi: 10.1088/1367-2630/aa6d3f.

[17] S.-Y. Lee . S.-Y. Lee, M. Widmann, T. Rendler, M. W. Doherty, T. M. Babinec, S. Yang, M. Eyer, P. Siyushev, B. J. M. Hausmann, M. Loncar, Z. Bodrog, A. Gali, N. B. Manson, H. Fedder, J. Wrachtrup, *Readout and control of a single nuclear spin with a metastable electron spin ancilla*, Nat. Nanotechnol. **8** 487 (2013), doi: 10.1038/nnano.2013.104.

[18] I. Aharonovich, S. Castelletto, D. A. Simpson, A. Stacey, J. McCallum, A. D. Greentree, S. Prawer, *Two-Level Ultrabright Single Photon Emission from Diamond Nanocrystals*, Nano Letters 9 (2009) 3191. doi: 10.1021/nl9014167

[19] D Gatto Monticone, P Traina, E Moreva, J Forneris, P Olivero, I P Degiovanni, F Taccetti, L Giuntini, G Brida, G Amato, M Genovese, *Native NIR-emitting single colour centres in CVD diamond*, New Journal of Physics 16 (2014) 053005. doi: 10.1088/1367-2630/16/5/053005.

[20] T Gaebel, I Popa, A Gruber, M Domhan, F Jelezko, J Wrachtrup, *Stable single-photon source in the near infrared*, New Journal of Physics 6 (2004) 98. doi:10.1088/1367-2630/6/1/098.

[21] J. Forneris, A. Tengattini, S. Ditalia Tchernij F. Picollo, A. Battiato, P. Traina, I.P. Degiovanni, E. Moreva, G. Brida, V. Grilj, N. Skukan, M. Jakšić, M. Genovese, P. Olivero, *Creation and characterization of He-related color centers in diamond*, J. Lumin. 179, 59 (2016). doi: 10.1016/j.jlumin.2016.06.039.

[22] A.M. Zaitsev, *Optical properties of Diamond* (Springer, New York, 2001).

[23] R. Brout, W. Visscher, *Suggested Experiment on Approximate Localized Modes in Crystals*, Phys. Rev. Letters 9 (1962) 54. doi: 10.1103/PhysRevLett.9.54.

[24] S. A. Solini, A. K. Ramnas, *Raman Spectrum of Diamond*, Phys. Rev. B 1 (1970) 1687. doi: 10.1103/PhysRevB.1.1687.

[25] A. Lohrmann, S. Pezzagna, I. Dobrinets, P. Spinicelli, V. Jacques, J.-F. Roch, J. Meijer, A.M. Zaitsev, *Diamond based light-emitting diode for visible single-photon emission at room temperature,* Appl. Phys. Lett. 99 (2011) 251106, http://dx.doi.org/10.1063/1.3670332.

[26] L. J. Rogers, K.D. Jahnke, M. W. Doherty, A. Dietrich, L. P. McGuinness, C. Müller, T. Teraji, H. Sumiya, J. Isoya, N. B. Manson, F. Jelezko, *Electronic structure of the negatively charged silicon-vacancy center in diamond*, Phys. Rev. B 89 (2014) 235101. doi: 10.1103/PhysRevB.89.235101.

[27] C. Hepp, T. Müller, V. Waselowski, J. N. Becker, B. Pingault, H. Sternschulte, D. Steinmüller-Nethl, A. Gali, J. R. Maze, M. Atatüre, C. Becher, *Electronic Structure of the Silicon Vacancy Color Center in Diamond*, Phys. Rev. Lett. 112 (2014) 036405. doi: 10.1103/PhysRevLett.112.036405.

[28] J.P. Goss, R. Jones, S.J.Breuer, P. R. Briddon, S. Öberg, *The Twelve-Line 1.682 eV Luminescence Center in Diamond and the Vacancy-Silicon Complex*, Phys. Rev. Lett. 77 (1996) 3041. doi.org/10.1103/PhysRevLett.77.3041.

[29] S.C. Kitson, P. Jonsson, J. G. Rarity, P.R. Tapster, *Intensity fluctuation spectroscopy of small numbers of dye molecules in a microcavity*, Physical Review A 58 (1998) 620. doi: 10.1103/PhysRevA.58.620.

[30] R. Brouri, A. Beveratos, J. P. Poizat, P. Grangier, *Photon antibunching in the fluorescence of individual color centers in diamond*, Opt. Lett. 25, 1294 (2000). doi: 10.1364/OL.25.001294.

[31] E. Neu, D. Steinmetz, J. Riedrich-Möller, S. Gsell, M. Fischer, M. Schreck. C. Becher, *Single photon emission from silicon-vacancy colour centres in chemical vapour deposition nano-diamonds on iridium*, New Journal of Physics 13 (2011) 025012. doi: 10.1088/1367-2630/13/2/025012.

[32] C. Kurtsiefer, S. Mayer, P. Zarda, H. Weinfurter, *Stable Solid-State Source of Single Photons*, Phys. Rev. Lett. 85 (2000) 290. doi: 10.1103/PhysRevLett.85.290.

[33] E. Wu, V. Jacques, H. Zeng, P. Grangier, F. Treussart, J.-F. Roch, *Narrow-band single-photon emission in the near infrared for quantum key distribution*, Opt. Express 14 (3) (2006) 1296. doi: 10.1364/OE.14.001296

[34] D. A. Simpson, E. Ampem-Lassen, B. C. Gibson, S. Trpkovski, F. M. Hossain, S. T. Huntington, A. D. Greentree, L. C. L. Hollenberg, S. Prawer, *A highly efficient two level diamond based single photon source*, Appl. Phys. Lett. 94 (2009) 203107. doi: 10.1063/1.3141450.

[35] A, Gruber, A. Dräbenstedt, C. Tietz, L. Fleury, J. Wrachtrup, C. von Borczyskowski, *Scanning confocal optical microscopy and magnetic resonance on single defect centers*. Science 276, 2012 (1997). doi: 10.1126/science.276.5321.2012

[36] J. Storteboom, P. Dolan, S. Castelletto, X. Li, M. Gu, *Lifetime investigation of single nitrogen vacancy centres in nanodiamonds*, Optics Express 23 (2015) 11327. doi: 10.1364/OE.23.011327.

[37] D. Steinmetz, E. Neu, J. Meijer, W. Bolse, C. Becher, *Single photon emitters based on Ni/Si related defects in single crystalline diamond*, Appl. Phys. B 102 (2011) 451−8. doi: 10.1364/OE.23.011327.

[38] P. Siyushev, V. Jacques, I. Aharonovich, F. Kaiser, T. Müller, L. Lombez, M. Atatüre, S. Castelletto, S. Prawer, F. Jelezko, J. Wrachtrup, *Low-temperature optical characterization of a near-infrared single-photon emitter in nanodiamonds*, New J. Phys. 11 (2009) 113029. 10.1088/1367-2630/11/11/113029.

[39] T. Müller, I. Aharonovich, L. Lombez, Y. Alaverdyan, A. N. Vamivakas, S. Castelletto, F. Jelezko, J. Wrachtrup, S. Prawer, M. Atatüre, *Wide-range electrical tunability of single-photon emission from chromium-based colour centres in diamond*, New J. Phys. 13 (2011) 075001. doi: 10.1088/1367-2630/13/7/075001.

[40] D. Gatto Monticone, F. Quercioli, R. Mercatelli, S. Soria, S. Borini, T. Poli, M. Vannoni, E. Vittone, P. Olivero, *Systematic study of defect-related quenching of NV luminescence in diamond with time-correlated single-photon counting spectroscopy*, Phys. Rev. B 88 (2013) 155201. doi: 10.1103/PhysRevB.88.155201.

[41] N.B.Manson, J.P.Harrison, *Photo-ionization of the nitrogen-vacancy center in diamond*. Diamond and Related Materials 14 (2005) 1705. doi:10.1016/j.diamond.2005.06.027.